\documentclass[12pt]{article}

\usepackage{amssymb}
\usepackage{color}
\usepackage{epsfig,amssymb,amsfonts,amsmath,graphicx,dsfont,cite,xfrac}
\usepackage{authblk}
\usepackage{subfigure}
\definecolor{mygray}{gray}{0.5}

\usepackage{cite}
\usepackage[colorlinks=true,linkcolor=blue,citecolor=red]{hyperref}
\newcommand{\be}{\begin{equation}}
\newcommand{\ee}{\end{equation}}
\newcommand{\bea}{\begin{eqnarray}}
\newcommand{\eea}{\end{eqnarray}}

\title{The Harmonic Oscillator in the Framework of Scale Relativity}

\author[${}$]{Moise Bonilla}
\author[${}$]{Oscar Rosas-Ortiz}

\affil[${}$]{\footnotesize Physics Department, Cinvestav, AP 14-740, 07000
M\'exico City, Mexico}

\date{}
\begin{document}

\maketitle

\begin{abstract}
The dynamical law obeyed by the one-dimensional physical systems in the scale relativity approach is reduced to a Riccati nonlinear differential equation. Applied to the harmonic oscillator potential, we show that such an approach permits the calculation of the solutions of the scale relativity problem in terms of the well known solutions of the Schr\"odinger equation for the harmonic oscillator. 
\end{abstract}

%%%%%%%%%%%%%%%%%%%%%%%%%%%%%%%%%%%%%%%%%%%%%%%

%---------------------------------------> Section
\section{Introduction}

Scale relativity is an emergent approach that aims to unify quantum physics and relativity theory \cite{Not11}. This is based on the notion that any measuring process in physics depends on the scale, so that no absolute measurements can be associated with any coordinate system. The latter would mean that the derivatives introduced by Newton must, at least, be revised. Introducing a new complex derivative which takes into account such scale dependence, the dynamical law of scale relativity is represented by a complex-differential equation with a term that encodes the nonclassicality of the system under study. Thus, by necessity, the velocity is a complex vector in such an approach. Then, besides the potential energy, the scale-dependent Hamiltonian includes a complex `kinetic energy' plus  a term that is proportional to the divergence of the complex velocity. If this last is different from zero, the Hamiltonian of scale relativity differs from the one of the conventional approaches even if the imaginary part of the velocity is zero. Thus, the divergence of the complex velocity is a measure of the nonclassicality of a given system in the scale relativity approach.

In this paper we solve the fundamental equation of scale relativity for the harmonic oscillator. As the latter is the simplest exactly solvable model in any of the physical theories, the results reported here are addressed to obtain a better understanding of the way in which scale relativity works. In particular we show that the solutions of the quantum problem are intimately connected with the ones of the scale relativity approach via the Riccati non-linear differential equation.

%---------------------------------->

\section{Basic equations}

In the simplest case, preserving the notion of Newtonian time, the space metric is continuous and non-differentiable everywhere \cite{Not11}. Then, the spatial displacements as well as the velocities are twice-valued. The complex velocity
\begin{equation}
\overrightarrow{\mathcal{V}} (\vec{x},t)  \equiv \vec{F}(\vec{x},t) - \mathrm{i} \vec{G}(\vec{x},t)
\end{equation}
is introduced to include the two velocities $\vec{v}_{\pm}$ since its real and imaginary parts are respectively given by $\vec{F}(\vec{x},t) \equiv \frac{\vec{v}_{+} + \vec{v}_{-}}{2}$ and $\vec{G}(\vec{x},t)  \equiv \frac{\vec{v}_{+} - \vec{v}_{-}}{2}$. In the Hamiltonian formulation \cite{Not16}, for a stationary system with definite energy $W$, the dynamical law is represented by the Riccati equation 
\begin{equation}
W = \frac{m}{2} \overrightarrow{\mathcal{V}}^{2} + \Phi
-   \mathrm{i}\frac{\hbar}{2} \big( \vec{\nabla} \cdot  \overrightarrow{\mathcal{V}} \big).
\label{rica1}
\end{equation}
For a particle of mass $m$ subjected to the one-dimensional potential $\Phi = \frac{1}{2} m \omega^{2} x^{2}$, the dynamical law (\ref{rica1}) takes the form	
\begin{equation}
\label{rica2}
\partial_{x}  \mathcal{V}   =  - \mathrm{i}\frac{m}{\hbar} \mathcal{V}^{2} + \mathrm{i}\frac{2}{\hbar} \Big( W - \Phi (x) \Big).
\end{equation}
As the complex velocity of a stationary state has no real part, we arrive at the expression
\begin{equation}
\label{rica3a}
-G^{\prime}  =  \frac{m}{\hbar} G^{2} + \frac{2}{\hbar} \Big( W - \frac{1}{2} m \omega^{2} x^{2} \Big).
\end{equation}
With the appropriate change of variables \cite{Bon16}, we rewrite (\ref{rica3a}) in dimensionless form
\begin{equation}
\label{rica3}
-g^{\prime} = g^{2} + \epsilon - \mathtt{x}^{2}.
\end{equation}
The latter nonlinear differential equation can be linearized by the logarithmic derivative $g(\mathtt{x}) = \frac{\mathrm{d}}{\mathrm{d}\mathtt{x}} \ln \xi (\mathtt{x})$. One obtains the Schr\"odinger-like equation
\begin{equation}
\label{schro1}
\mathrm{H} \xi = \Big( -\frac{\mathrm{d}^{2}}{\mathrm{d}\mathtt{x}} + \mathtt{x}^{2} \Big) \xi = \epsilon \, \xi,
\end{equation}	
where $\mathrm{H} = - \frac{\mathrm{d}^{2}}{\mathrm{d}\mathtt{x}^{2}} + \mathtt{x}^{2}$  is the related Hamiltonian and $\epsilon$ corresponds to the energy eigenvalue. That is, (\ref{schro1}) is the eigenvalue equation associated with the (mathematical) harmonic oscillator $\Phi=\mathtt{x}^{2}$. We are going to take full advantage of the connection between (\ref{rica3}) and (\ref{schro1}) to solve the dynamical problem of the scale relativity for the harmonic oscillator. That is, we shall obtain the solutions of (\ref{rica3}) by solving the eigenvalue problem (\ref{schro1}).

%---------------------------------->	
\section{Solution of the problem}

It is well known that the operators
\begin{equation}
\label{aes}
\mathtt{a}  =   \frac{\mathrm{d}}{\mathrm{d}\mathtt{x}} + \mathtt{x}, \quad \mathtt{a}^{\dagger}  =   -\frac{\mathrm{d}}{\mathrm{d}\mathtt{x}} + \mathtt{x}, \quad \big(\mathtt{a}^{\dagger}\big)^{\dagger}  =  \mathtt{a},
\end{equation}
satisfy the Heisenberg algebra
\begin{equation}
[\mathtt{a}, \mathtt{a}^{\dagger}] = 2\mathrm{I},  \quad 
[\mathrm{H}, \mathtt{a}] = - 2 \mathtt{a},  \quad  [\mathrm{H}, \mathtt{a}^{\dagger}] =   2 \mathtt{a}^{\dagger},
\label{algebra}
\end{equation}
and factorize the Hamiltonian $\mathrm{H}$ as follows
\begin{equation}
\label{factor}
\mathtt{a}\mathtt{a}^{\dagger}  =  -\frac{\mathrm{d}^{2}}{\mathrm{d}\mathtt{x}^{2}} + \mathtt{x}^{2} + 1 =  -\frac{\mathrm{d}^{2}}{\mathrm{d}\mathtt{x}^{2}} + \mathtt{x}^{2} - \epsilon \equiv  \mathrm{H}-\epsilon.
\end{equation}
The simplest form of solving the eigenvalue equation (\ref{schro1}) considers the factorization (\ref{factor}) and an extremal function $\phi_{0}(\mathtt{x})$ which is annihilated by the operator $\mathtt{a}$. That is
\begin{equation}
\mathtt{a}\phi_{0}(\mathtt{x})=0 \quad \Rightarrow \quad \mathrm{H}\phi_{0}(\mathtt{x})=\phi_{0}(\mathtt{x}).
\end{equation}
If $\phi_{0}(\mathtt{x})$ is of finite norm then it is a normalizable eigenfunction of $\mathrm{H}$ with eigenvalue $E_{0}=1$. Using (\ref{aes}) it is immediate to obtain
\begin{equation}
\phi_{0}(\mathtt{x})=C_{0}e^{-\frac{\mathtt{x}^{2}}{2}}.
\end{equation}
Clearly $\phi_{0}(\mathtt{x})\in \mathrm{L}^{2}(\mathbb{R})$, so that the integration constant  $C_{0}$ is fixed by normalization. On the other hand, from the oscillation theorem we know that there is no square-integrable solution belonging to the eigenvalue $E<E_{0}$ since $\phi_{0}(\mathtt{x})$ is free of nodes. Therefore, $\phi_{0}(\mathtt{x})$ is the wave function of the ground energy of the oscillator. 

Iterating $n$-times the action of $\mathtt{a}^{\dagger}$ on $\phi_{0}(\mathtt{x})$ one obtains 
\begin{equation}
\phi_{n}(\mathtt{x}) = \frac{C_n}{\sqrt{ \sqrt{\pi} \, 2^{n}n!}}\big( \mathtt{a}^{\dagger} \big)^{n} \phi_{0}(\mathtt{x}),
\end{equation}
which is the wave function of the state associated to the energy $E_{n}= E_{0}+2n$. After the straightforward calculation we have
\begin{equation}
\phi_{n}(\mathtt{x}) = C_{n} \exp \Big( - \frac{\mathtt{x}^{2}}{2} \Big) \mathrm{H}_{n} (\mathtt{x}),
\label{fisol}
\end{equation}
with $C_n$ the normalization constant and $\mathrm{H}_{n}(\mathtt{x})$ standing for the Hermite polynomials \cite{Abr70}.

A first set of solutions to the Riccati equation (\ref{rica3}) are given by the logarithmic derivative of the wave functions
\begin{equation}
\label{ges}
g_n(\mathtt{x}) = \frac{\mathrm{d}}{\mathrm{d}\mathtt{x}} \ln \phi_{n} (\mathtt{x}) = -\mathtt{x} + \frac{2n\mathrm{H_{n-1}(\mathtt{x})}}{H_{n}(\mathtt{x})} = \mathtt{x} - \frac{\mathrm{H_{n+1}(\mathtt{x})}}{H_{n}(\mathtt{x})}.
\end{equation}
That is,
\begin{equation}
g_0(\mathtt{x}) = -\mathtt{x}, \quad  g_1(\mathtt{x}) = -\mathtt{x} + \frac{1}{\mathtt{x}}, \quad g_2(\mathtt{x}) = -\mathtt{x} + \frac{4\mathtt{x}}{2\mathtt{x}^{2}-1}, \quad \ldots
\end{equation}
Additionally, there is a solution of (\ref{rica3}) which cannot be obtained from the functions (\ref{fisol}). Namely, for $\epsilon=-1$ the simplest solution is $g_{-1}(\mathtt{x})=\mathtt{x}$. However, such a function gives rise to a solution of (\ref{schro1}) that is not normalizable 
\begin{equation}
\phi_{-1}(\mathtt{x})= \exp \Big( \frac{\mathtt{x}^{2}}{2} \Big) \quad \Rightarrow \quad \mathrm{H} \phi_{-1}(\mathtt{x}) = -\phi_{-1}(\mathtt{x}).
\label{missing}
\end{equation}
The same holds for the solutions of the Riccati equation (\ref{rica3}) for any other $\epsilon < E_0 =1$. In this form, among the solutions of the scale relativity equation (\ref{rica3}), only the $g$-functions obeying the transformation (\ref{ges}) admit an interpretation in the quantum approaches \cite{Bon16}. We may write
\begin{equation}
g_{n}^{\prime}(\mathtt{x}) + g_{n}^{2}(\mathtt{x}) = \mathtt{x}^{2} - 2n -1, \quad n=0,1,2,\ldots
\label{ricafin}
\end{equation}
as the equation in scale relativity that defines physical solutions in the quantum picture for the harmonic oscillator.  

%%%%%%%%%%%%%%%%%%%%%%
\begin{figure}[htb]
\centering
\subfigure[]{\includegraphics[width=0.23\textwidth]{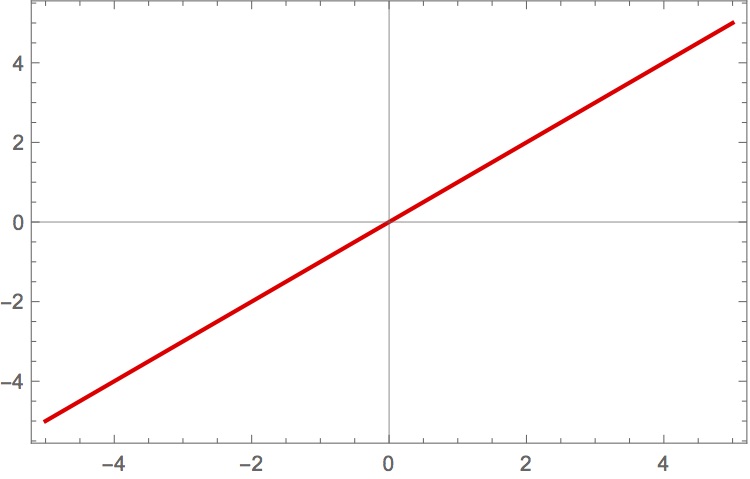}} 
\hspace{1ex}
\subfigure[]{\includegraphics[width=0.23\textwidth]{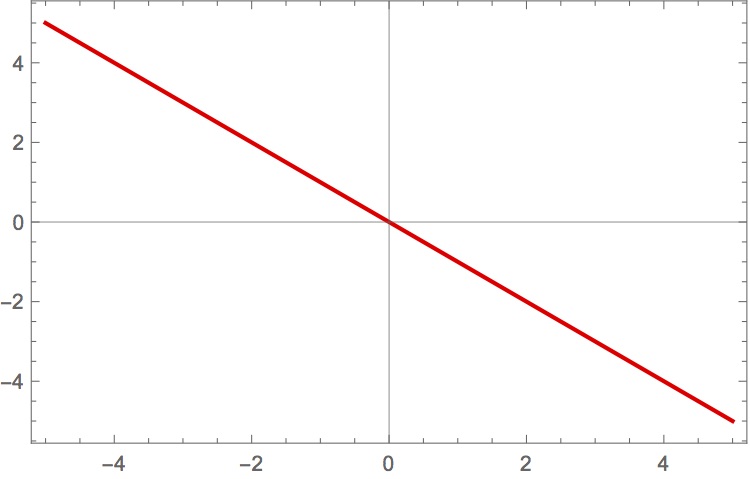}} 
\hspace{1ex}
\subfigure[]{\includegraphics[width=0.23\textwidth]{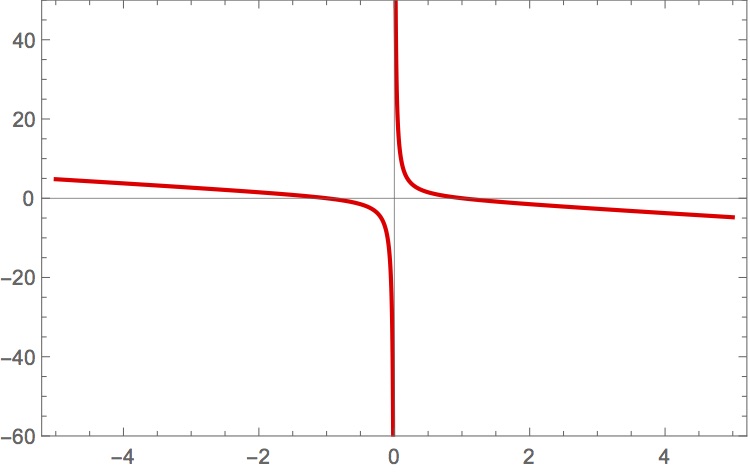}} 
\hspace{1ex}
\subfigure[]{\includegraphics[width=0.23\textwidth]{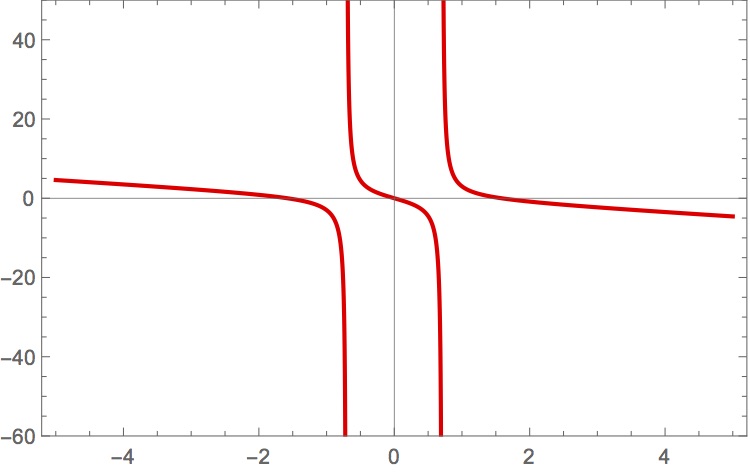}}

\centering
\subfigure[]{\includegraphics[width=0.23\textwidth]{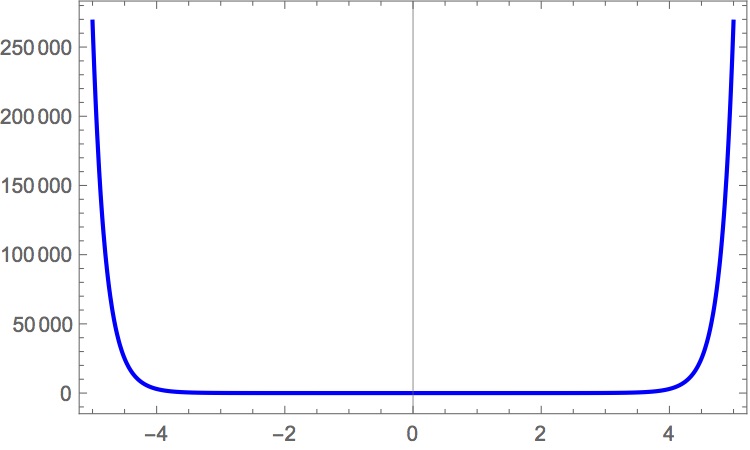}} 
\hspace{1ex}
\subfigure[]{\includegraphics[width=0.23\textwidth]{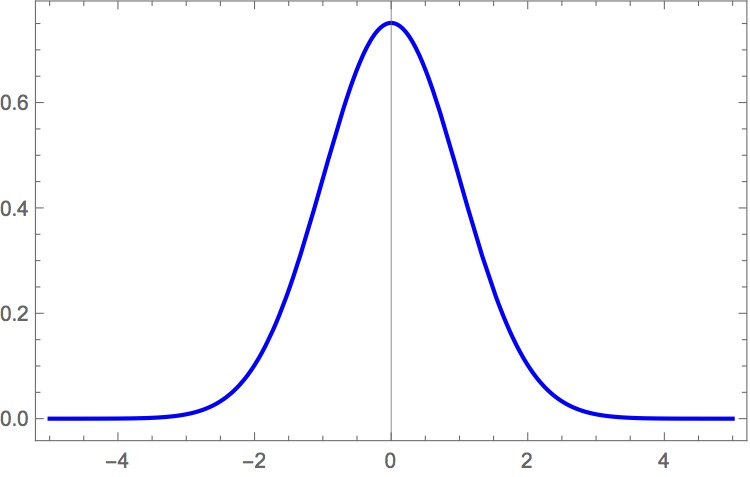}} 
\hspace{1ex}
\subfigure[]{\includegraphics[width=0.23\textwidth]{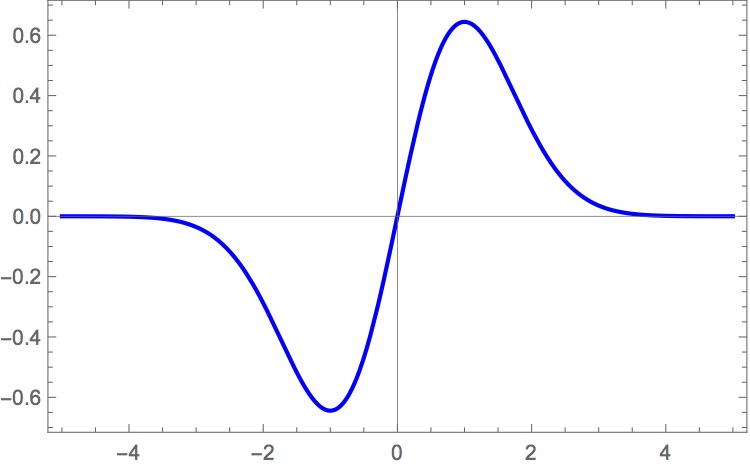}} 
\hspace{1ex}
\subfigure[]{\includegraphics[width=0.23\textwidth]{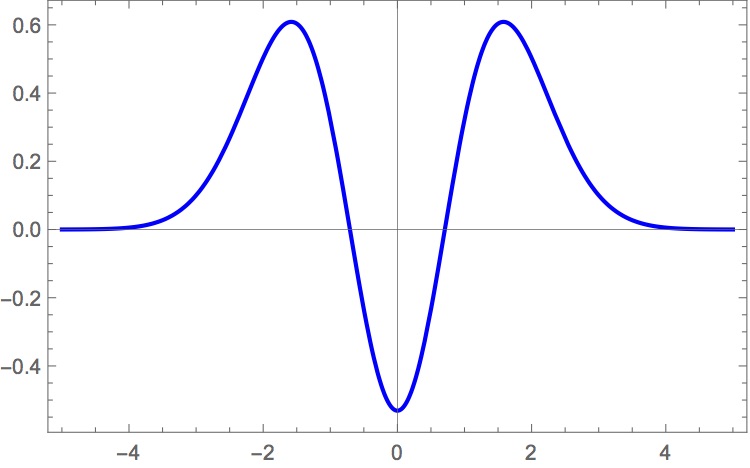}}

\caption{\footnotesize 
The upper and lower rows include some solutions of the  Riccati equation (\ref{rica3}) and the Schr\"odinger equation (\ref{schro1}) respectively. From left to right the columns correspond to $\epsilon = -1$, $\epsilon=E_0$, $\epsilon= E_1$, and $\epsilon = E_2$, with $E_n=2n+1$, $n=0,1,2,\ldots$,  the energy eigenvalues of the quantum harmonic oscillator. Notice that $g_{-1}$ is connected with the function $\phi_{-1}$ which is not square-integrable, see Eq.~(\ref{missing}).
}
\label{fig1}
\end{figure}
%%%%%%%%%%%%%%%%%%%%%%

In the panel of Fig.~\ref{fig1} we show some of the solutions $g_n$ of (\ref{rica3}) as well as the corresponding solutions $\phi_n$ of (\ref{schro1}). Note that, in all the cases, the $g$-functions diverge as $\vert x \vert \rightarrow \infty$. Besides, we can appreciate that the singularities of $g_1$ and $g_2$ are associated with the nodes of $\phi_1$ and $\phi_2$ respectively. In turn, the function $g_0$ has no singularities since the ground state wave function $\phi_0$ is free of nodes. Interestingly, although $g_{-1}$ is free of singularities and diverges at $x=\pm \infty$, it is  associated with the non square-integrable function $\phi_{-1}$. The latter because the slopes of $g_{-1}$ and $g_0$ have opposite sign. Another interesting property of the solutions of (\ref{ricafin}) is that their zeros are in connection with the local maxima and minima of $\phi_n$. On the other hand, it is well known that the $\phi_n$ are nonclassical states for $n \geq 1$ since their $P$-function is as singular as the derivatives of the delta distribution \cite{Gla07}. In contrast, the ground state $\phi_0$ is classical because its $P$-function is equal to $\delta (x)$. The relationship between the singularities of $g_n$ and the nonclassicality of $\phi_n$ is in progress and will be reported elsewhere.

%---------------------------------->
\section{Concluding remarks}

We have shown that the fundamental equation of the scale relativity for the harmonic oscillator can be reduced to a Riccati nonlinear differential equation. Using the well known relationship between the Riccati and the Schr\"odinger equations we have solved the scale relativity problem in terms of the oscillator quantum eigenfunctions. In contrast with other works \cite{Not16}, we require no numerical approximations to justify the form of the solutions. We have found some interesting relationships between the singularities of the scale relativity solutions and the nodes of the quantum wave functions. Further progress will be reported elsewhere. 

%----------------------------->
\section*{Acknowledgment}
M.B. acknowledges the funding received through a CONACyT scholarship.

%-------------------------------------->

\end{document}